\documentclass{article}

\usepackage{spconf,amsmath,graphicx}
\usepackage{xcolor}
\usepackage{cite}
\usepackage{url}
\usepackage{amssymb}
\usepackage[ruled,vlined]{algorithm2e}
\usepackage{flushend}

\def\s{{\mathbf{S}_1}}
\def\st{{\mathbf{S}_1^T}}
\def\se{{\mathbf{S}_2}}
\def\sst{{\mathbf{S}_2^T}}
\def\frf{{\mathbf{F}_{RF}}}
\def\frfh{{\mathbf{F}_{RF}^H}}

\title{Learning to Select for MIMO Radar based on Hybrid Analog-Digital Beamforming}
\name{$\text{Zhaoyi Xu}^{1}$\thanks{Work supported in part by NSF under grant ECSS-2033433, and in part by the Marie Sk{\l}odowska-Curie Individual Fellowship under grant 793345.}, $\text{Fan Liu}^{2}$, $\text{Konstantinos Diamantaras}^{3}$, $\text{Christos Masouros}^{2}$,  and $\text{Athina Petropulu}^{1}$}
\address{$^{1}$Dept. of Electrical and Computer Engineering, Rutgers University,\\
$^{2}$Dept. of Electronic and Electrical  Engineering, University College London\\
$^{3}$Dept. of Information and Electronic Eng., International Hellenic University,  Thessaloniki, Greece}

\begin{document}

\maketitle

\begin{abstract}
In this paper, we propose an energy-efficient radar beampattern design framework for Millimeter Wave (mmWave) massive multi-input multi-output (mMIMO) systems,  equipped with a hybrid analog-digital (HAD) beamforming structure. Aiming to reduce the power consumption and hardware cost of the mMIMO system, we employ a learning  approach to synthesize the probing beampattern based on a small number of RF chains and antennas. By leveraging a combination of softmax neural networks, the proposed solution is able to achieve a desirable  beampattern with high accuracy while incurring low cost.
\end{abstract}
\section{Introduction}
Sensing is viewed as an essential feature in  next-generation wireless communication applications \cite{6G_whitepaper}, such as  vehicular networks, WLAN indoor positioning, and unmanned aerial vehicle (UAV) networks \cite{Fan2020}. Indeed, in all  those  scenarios, sensing and communication are a pair of intertwined functionalities, often required to be carried out simultaneously for the purpose of increasing the spectral efficiency and reducing costs.

In order to promote both high-throughput communication and high-accuracy sensing performance, millimeter Wave (mmWave) signaling and  massive multi-input multi-output (mMIMO)  have emerged as two promising approaches \cite{6515173,7400949,5595728}. The large bandwidth available at the mmWave spectrum provides not only  Moreover, the large-scale antenna array can compensate for the path-loss of the mmWave signals by formulating ``pencil-like" beams towards the communication users. At the same time, a large-scale antenna array offers enhanced performance in terms of the angular resolution for radar sensing. However, fully-digital mMIMO systems require as many RF chains as antenna elements. This requirement translates into high power consumption and hardware cost, which limit the applicability of fully-digital mMIMO in a practical setting, especially when the antennas and RF chains are operated in the mmWave band. \\
\indent The hybrid analog-digital (HAD) beamforming structure is a low-cost solution for tackling the above issues \cite{8030501} while  reaping the performance gains of both mMIMO and mmWave signalling. A HAD architecture comprises a small number of RF chains, which are connected to a large number of antennas through a network of phase shifters. While the HAD beamforming for communication has already been well-studied\cite{DL_hybrid_mmwave,hojatian2020unsupervised,NN_hybrid_mmwave,hybrid_selection,8778669}, its application towards radar sensing remains to be explored. To this end,  previous research efforts have focused on the design of phased-MIMO radar, which was proposed as a tradeoff between the phased-array and MIMO radars \cite{5419124}. However, due to the exponentially increasing complexity and energy consumption in terms of both antennas and RF chains,  the state-of-the-art research on phased-MIMO radar is restricted to small-scale antenna arrays \cite{5419124,6104178,6376087}, and thus cannot take advantage of the mMIMO capabilities. To address this issue, it is necessary to exploit a limited number of RF chains and/or antennas instead of using the full HAD array. Again, to the best of our knowledge, the literature on  antenna/RF chain selection  for phased-MIMO/HAD radar is rather sparse.
\\\indent To further reduce the cost and improve energy efficiency of the conventional phased-MIMO radar \cite{5419124}, we propose a novel approach for optimally selecting a small number of RF chains and/or antennas from a dense hybrid analog-digital array, along with optimally designing the phase shifter network matrix and the beamforming matrix, so that the corresponding probing beampattern is close to that of a fully populated HAD structure. 
The optimization problem is solved by modifying the softmax learning approach \textit{learn to select} (L2S) in \cite{SPL} where the selection of antennas is modeled by softmax neural networks. The proposed L2S is effective in formulating any desirable radar beampattern and can scale to a large number of RF chains and/or antennas to select from which is crucial to massive array.
\\\indent While machine learning for antenna selection has been investigated in  \cite{Joung-2016,hybrid_selection,Vu2019machine}, the problem in those works was treated as a classification problem. However, the combinatorial explosion problem renders those methods impractical even in cases with a  moderate number of antennas. On the other hand, 
L2S in \cite{SPL} can be efficiently scaled to larger problems as it avoids the combinatorial explosion of the selection problem. It also offers a flexible array design framework as the selection problem can be easily formulated for any metric. 
For clarity, we note here that \cite{SPL} considers a sparse array design problem where only one selection matrix is considered. In contrast, the problem considered in this paper involves two selection matrices, i.e., a phase-shifter network matrix with unit modulus, and a beamforming matrix. Both matrices  are parameters in the optimization problem, thus 
providing more degrees of freedom for approximating the desired beampattern while incurring lower cost. 

\section{Problem formulation}
\label{sec:format}

Let us consider a massive MIMO system  equipped with $N_t$ antennas and $N_{RF}$ RF chains. The antennas  formulate a uniform linear array, with spacing between adjacent antennas denoted by $d$. In the fully digital MIMO system, we  have $N_{RF} = N_t$, which large huge costs when a large number of  antennas are needed, especially in the case of RF chains operating in the mmWave band. To tackle this issue, we consider a HAD structure, which employs a smaller number of RF chains, i.e., $N_{RF} \le N_t$, and where each RF chain is connected to all $N_t$ antennas via a phase-shifter network.


The phase-shifter network between antennas and RF chains can be modeled as a matrix ${{\mathbf{F}}_{RF}} \in {\mathbb{C}^{N_t \times {N_{RF}}}}$, where all the entries in ${{\mathbf{F}}_{RF}}$ have constant modulus, i.e., $\left| {{{\mathbf{F}}_{RF}}\left( {i,j} \right)} \right| = 1,\forall i,j$.
We assume that the antennas transmit narrow-band signals with carrier wavelength $\lambda$.
The array output at angle $\theta$ is
\begin{equation}
    y(t;\theta) = \mathbf{a}(\theta)^H \mathbf{v}(t),
\end{equation}
where $\mathbf{a}(\theta)$ is the steering vector at direction $\theta$, and $\mathbf{v}(t) \in \mathbb{C}^{N_t}$ is the transmit array snapshot at time $t$.
Let
\begin{equation}
    \mathbf{v}(t) = \frf\mathbf{Q}\mathbf{e}(t)
\end{equation}
with $\mathbf{Q}\in \mathbb{C}^{N_{RF}\times N_{RF}}$  a baseband precoding matrix, and $\mathbf{e}(t) \in \mathbb{C}^{N_{RF}\times 1}$  a white signal vector with zero-mean and unit identity covariance matrix.
The array output vector at $K$ different angles is
\begin{eqnarray}
   \mathbf{y} &\triangleq& [y(t;\theta_1),\dots,y(t;\theta_K)]^T \\
   &=& \mathbf{A}^H \mathbf{v}(t) = \mathbf{A}^H\frf\mathbf{Q}\mathbf{e}(t)
\end{eqnarray}
 with
$\mathbf{A}=[\mathbf{a}(\theta_1), \dots, \mathbf{a}(\theta_K)] \in \mathbb{C}^{N_t\times K}$ being the steering matrix.
In order to achieve a lower power consumption and hardware cost, we want to select $M_{RF}$ RF chains ($M_{RF}<N_{RF}$) and $M_{t}$ array antennas ($M_{t}<N_t$), so that the HAD system  approximates the desired power pattern at $K$ angles.
To proceed, let us first select $M_{RF}$ out of $N_{RF}$ RF chains by multiplying ${{\mathbf{F}}_{RF}}$ with a selection matrix ${{\mathbf{S}}_1} \in {\mathbb{C}^{{M_{RF}} \times N_{RF}}}$, where all the elements in $\s$ are zero, except for exactly one element per row which is equal to one. The transmitted snapshot from the selected RF chains can be expressed as
\begin{equation}
    \mathbf{v}_s(t) = \frf\st\s\mathbf{Q}\mathbf{e}(t)
\end{equation}
\textcolor{black}{Each column in $\mathbf{S}_1$ contains at most one element that is equal to one so that we do not choose the same RF chain twice.}
%


\textcolor{black}{In general, if we select RF chains of indices $l_1,\dots,l_{M_{RF}}$ then the elements of $\s$ will be $ s_{1_{ij}} =1$ for $l_i = j$, and $0$ otherwise.
}
Correspondingly,  matrix $\st\s$ will be an $N_{RF}\times N_{RF}$ diagonal matrix where the diagonal entries are one if the corresponding RF chains are active and zero otherwise. Similarly, another selection matrix $\se \in \mathbb{C}^{M_{t} \times N_t}$ can be introduced to select $M_{t}$ out of $N_t$ antennas.

The output of the sparse array can be expressed as
\begin{eqnarray}
& &\mathbf{y}_s(t) \triangleq [y_s(t;\theta_1), \dots, y_s(t;\theta_K)]^T \\
&=& \mathbf{A}^H\sst\se \mathbf{v}_s(t)
= \mathbf{A}^H\sst\se\frf\st\s\mathbf{Q}\mathbf{e}(t)
\end{eqnarray}

Let $p_i = p(\theta_i)$ be the desirable signal power at direction $\theta_i$, so that the desired beampattern vector is $\mathbf{p}=[p_1,\dots,p_K]^T$.
The sparse array output power at $\theta_i$ is
\begin{eqnarray}
    \hat{p}_i &=& \mathbb{E}\{y_s(t;\theta_i)^* y_s(t;\theta_i)\} \\
    &=&
     \mathbf{a}^H(\theta_i) \sst \se\frf\st\s\mathbf{Q}\nonumber\\
    &&\times\mathbf{Q}^H\st\s\frfh\sst \se\mathbf{a}(\theta_i)
\end{eqnarray}
The goal is to find the  selection matrices $\s$, $\se$, the mapping matrix $\frf$ and the precoding matrix $\mathbf{Q}$ that minimize the beam-pattern error, i.e., 
\begin{equation}
    \begin{aligned}
    \min_{\s,\se,\frf,\mathbf{Q}} \quad 
    &\sum_{i=1}^K (p_i - \hat{p}_i)^2 \\
     \textrm{s.t.} \qquad\quad & |\frf(i,j)|^2 = 1,\forall i,j; \nonumber \\
    & \s\st = \mathbf{I}_{M_{RF}}; \quad 
     \se\sst = \mathbf{I}_{M_{t}}
    \end{aligned}
\end{equation}
where $\sum_{i=1}^K (p_i - \hat{p}_i)^2 = \|\mathbf{p} - \text{diag}\{\mathbf{A}^H  \sst \se\frf\st\s\times\mathbf{Q}\mathbf{Q}^H\st\s\frfh\sst \se \mathbf{A}\}\|^2$ and $\mathbf{I}_M$ is an $M\times M$ identity matrix.

\section{Softmax Co-design}
We propose to use the  learning approach in \cite{SPL} for the co-design of  $\s$, $\se$, $\frf$ and $\mathbf{Q}$. Let the loss function be
\begin{align}
    \mathcal{L}(\s,\se,\frf,\mathbf{Q}) =&  \|\mathbf{p} - \text{diag}\{\mathbf{A}^H  \sst\nonumber \se\frf\st\s\nonumber\\
    &\times\mathbf{Q}\mathbf{Q}^H\st\s\frfh\sst \se \mathbf{A}\}\|^2.
    \label{loss}
\end{align}
Each row of selection matrices $\s$ and $\se$ can be modeled by a separate softmax neural network \cite{bishop2006pattern}. Taking the RF chain selection matrix $\s$ as an example, the outputs of the $m$-th network will be
\begin{equation}
    s_{m,i} = \frac{\exp(\mathbf{w}_i^T \mathbf{x} + b_{i})
    }{\sum_{j=1}^{N_{RF}} \exp(\mathbf{w}_j^T \mathbf{x} + b_{j})
    }, ~~~~i=1,\dots,N_{RF}
\end{equation}
where $\mathbf{w}_i$, $b_i$ are respectively the weights and biases, and $\mathbf{x}$ is the input.
Note that
$
0 \leq s_{m,j}\leq 1
$
and
\begin{equation}
    \sum_{j=1}^{N_{RF}}s_{m,j} = 1.
    \label{eq:sum_s}
\end{equation}
Essentially, $s_{m,i}$ represents the probability that RF chain $i$ will be our $m$-th selected RF chain.

Since the selection matrix does not depend on time $t$, the input $\mathbf{x}$ should be constant, and thus, the constant value $b_i' = \mathbf{w}_i^T \mathbf{x}$ can be merged into the bias term $b_i$.
Without loss of generality, such a model is equivalent to a softmax model with $\mathbf{x}=0$, where the only trainable parameters are the biases.

The approximate selection matrix, $\mathbf{\hat{S}}_1$, is formed based on the outputs $\mathbf{s}_m = [s_{m,1},\dots,s_{m,N_{RF}}]$ of all the softmax models as its rows.
Clearly, $\mathbf{\hat S}_1$ will be a soft selection matrix since the values $s_{m,i}$ range between 0 and 1.
By the end of the training, the matrix should converge very close to hard binary values so the approximation will be successful.

In order to formulate the cost function we individually express $y_s(t;\theta_k)$ in terms of real and imaginary parts, to facilitate the machine learning optimization which is based on real numbers.

The average output power at angle $\theta_k$ is 
\begin{eqnarray}
    \tilde{p}_k &=& \frac{1}{T} \sum_{t=1}^T y_s^*(t;\theta_k) y_s(t;\theta_k)
\end{eqnarray}
and the  beampattern error with respect to $p_k$ is
\begin{equation}
    \mathcal{\tilde L} = \sum_{k=1}^K \gamma_k (p_k - \tilde{p}_k)^2
\end{equation}
where $\gamma_k$ is the importance weight assigned to the angle $\theta_k$.

In order to achieve a realistic solution, the softmax models must produce hard binary values. The following constraint enforces this requirement:
\begin{equation}
    \sum_{i=1} s_{m,i}^2 = 1,
    \forall m.
    \label{eq:constraint_1}
\end{equation}
Indeed, \eqref{eq:constraint_1} holds iff $s_{mi} \in \{0, 1\}$.
The `if' part of this statement is obvious.
The `only if' part comes readily from \eqref{eq:sum_s} since
\begin{align*}
\Bigl[\sum_{i=1} s_{mi}\Bigr]^2 - \sum_{i=1} s_{mi}^2 = 0
\Rightarrow 2\sum_{i\neq j} s_{mi} s_{mj} = 0
\end{align*}
implying that at most one element of $\mathbf{s}_m$ can be equal to 1 and all other elements must be equal to 0.
Combined with \eqref{eq:sum_s} this means that \textit{exactly} one element of $\mathbf{s}_m$ is equal to 1 and all other elements are equal to 0.

We also need to impose another constraint since the same RF chain or antenna can not be selected more than once, i.e.
\[
    s_{m,i} = 1 \Rightarrow s_{n,i} = 0,
    \forall n\neq m
\]
If $s_{m,i} \in \{0,1\}$ then the above constraint is equivalent to
\begin{equation}
    \mathbf{s}_m^T \mathbf{s}_n = 0.
    \label{eq:constraint_2}
\end{equation}
Combining \eqref{eq:constraint_1} and \eqref{eq:constraint_2} it follows that $\mathbf{\hat S}_1\mathbf{\hat S}_1^T$ must be equal to the identity matrix $\mathbf{I}_{M_{RF}}$.
Based on the power gain error and the selection matrix structure above, we formulate the following loss function: 
\begin{equation}
    \mathcal{L}_0(\mathbf{b}_1,\mathbf{b}_2,\mathbf{F}_{RF},\mathbf{Q})
    = \mathcal{\tilde L} + \alpha_1 \|\mathbf{\hat S}_1\mathbf{\hat S}_1^T - \mathbf{I}\|_F^2
    + \alpha_2\|\mathbf{\hat S}_2\mathbf{\hat S}_2^T - \mathbf{I}\|_F^2.
    \label{eq:learning_loss}
\end{equation}
where $\|\cdot\|_F$ denotes the matrix Frobenius norm, and $\alpha_1$ and $\alpha_2$ are cost parameters which reflect the relative importance of the latter constraint with respect to the desired beam-pattern error.

\subsection{Learning to select RF chains and antennas}

There are four sets of parameters to be trained:
(i) the biases $\mathbf{b}_1$ to approximate the selection on RF chains , (ii) the biases $\mathbf{b}_2$ to approximate the selection on antennas (assuming $\mathbf{x}=0$), (iii) the covariance shaping matrix $\mathbf{Q}$ and (iv) the phase-shifter network matrix $\frf$.
We propose a four-stage optimization approach, by alternating between optimizing over one set of parameters and fixing others.

The algorithm runs for $N_{epoch}$ learning epochs and each alternating stage runs for a small number of steps $N_{step}$.
The proposed  scheme  is shown in Algorithm \ref{alg:learn_to_select}.
One can improve the speed of convergence by using other optimizers instead of gradient descent. In the simulations shown next we used the Adam optimizer \cite{Kingma2014adam}.

\begin{algorithm}[h]
\SetAlgoVlined
\DontPrintSemicolon
\For{epoch=$1$ \KwTo $N_{epochs}$}{
    Fix $\mathbf{Q}$, $\frf$, $\mathbf{b}_2$ and optimize $\mathcal{L}_0$ w.r.t. $\mathbf{b}_1$:\;
    \For{step=$1$ \KwTo ~$N_{steps}$}{
      Update $\mathbf{b}_1$\;
    }
    Fix $\mathbf{b}_1$, $\mathbf{b}_2$, $\mathbf{Q}$ and optimize $\mathcal{L}_0$ w.r.t. $\frf$:\;
    \For{step=$1$ \KwTo $N_{steps}$}{
      Update $\frf$\;
    }
    Fix $\frf$, $\mathbf{b}_1$, $\mathbf{Q}$ and optimize $\mathcal{L}_0$ w.r.t. $\mathbf{b}_2$:\;
    \For{step=$1$ \KwTo ~$N_{steps}$}{
      Update $\mathbf{b}_2$\;
    }
    Fix $\mathbf{b}_1$, $\mathbf{b}_2$, $\frf$ and optimize $\mathcal{L}_0$ w.r.t. $\mathbf{Q}$:\;
    \For{step=$1$ \KwTo $N_{steps}$}{
      Update $\mathbf{Q}$\;
    }
}
\caption{Learn to select.}
\label{alg:learn_to_select}
\end{algorithm}

\section{Simulation results}
Here, we demonstrate  the performance and flexibility of the proposed method. In all experiments, a flat weight is used, i.e., $\gamma_k = 1,\forall k$ and the antennas are spaced by half of wavelength. 
We used the Adam stochastic optimization procedure with different learning rates and $N_{epoch}=400$ epochs of training. In each epoch $N_{step}=10$ steps are executed. The training data are 
\text{i.i.d.}
jointly complex  Gaussian with zero mean and variance $1$. The length of input data $T$ should always be larger than the maximum number of $M_{RF}$ and $M_{t}$ to ensure the functionality of the model. 

Our first experiment is designed to select a small number of RF chains to reduce the system cost. 
The desirable beam power profile  equals to $1$ over the angle ranges $[-27,-23]$ degrees and $[28,32]$ degrees, and is zero otherwise.
 The number of antennas is $N_t=128$. The learning rate is set to $\beta = 0.04$, while the parameter $\alpha_1$, used in \eqref{eq:learning_loss}, changes between learning epochs;  it starts from  $\alpha_{init} = 3200$, and  linearly increases to  $\alpha_{final}= 16000$ at the final epoch. The $\alpha$ weights the importance of a proper selection matrix during the learning process.

\begin{figure}
{\small
    \centering
    \includegraphics[width = 8cm]{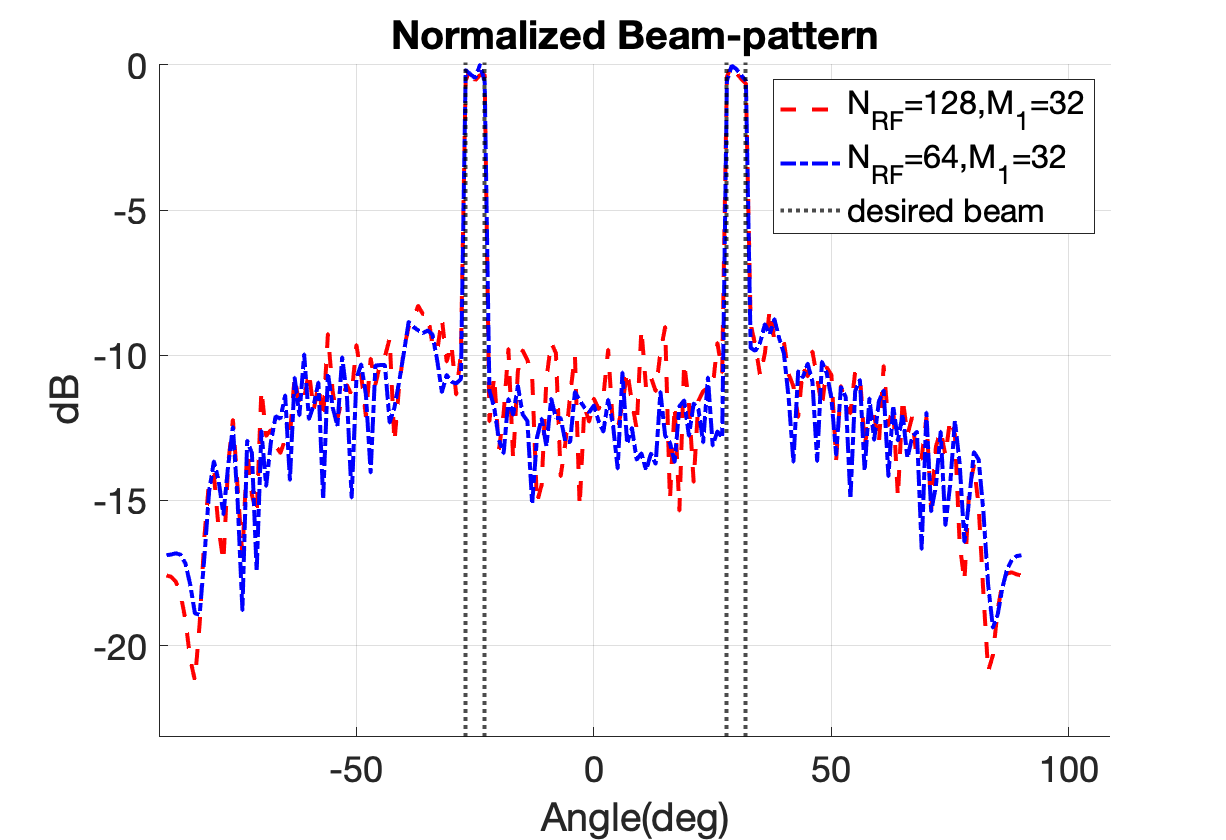}
    \caption{Beampatterm based on selecting $M_{RF} =32$  out of $N_{RF}=64$ (blue), or  $128$ (red) RF chains. $N_t = 128$.}
    \label{fig1}
    }
\end{figure}

Fig. \ref{fig1}, shows the designed beampattern,  when selecting $M_{RF} = 32$ out of $N_{RF} = 64$ or $N_{RF} = 128$ RF chains, \textcolor{black}{which are typical antenna numbers considered in mMIMO systems}. One can see that the matching to the desirable beampattern is pretty good.
\textcolor{black}{For this example, classification-based machine learning methods  would have to choose the best class out of ${{64}\choose{32}} > 1.8\times 10^{18}$ classes, which  is a task that would require a prohibitively long time to compute.}

\begin{figure}
    \centering
    \includegraphics[width = 8cm]{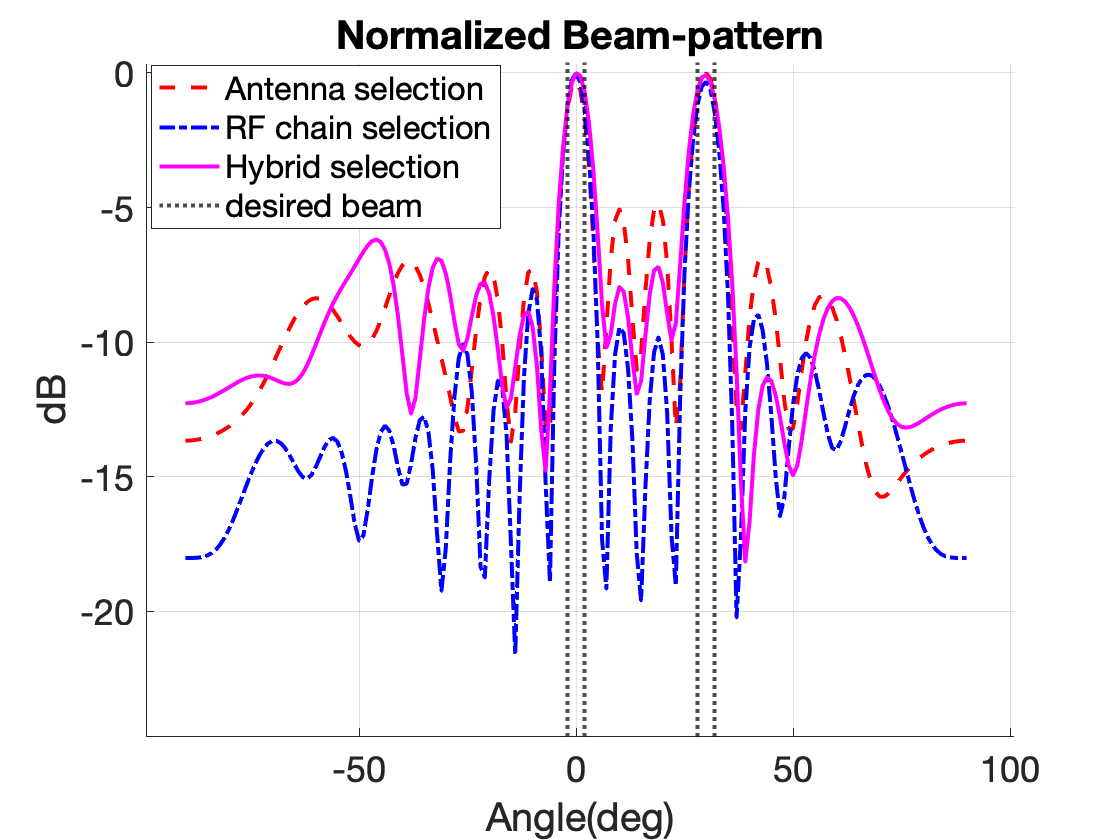}
    \caption{Selection in HAD array compared with selection in the full digital array.}
    \label{fig2}
\end{figure}

In the second experiment, three different selection choices are tested: (i) select  antennas only, (ii) select RF chains only, and (iii) select both RF chains and antennas. 
There are  $N_t=64$ antennas and  $N_{RF}=32$ RF chains, among which we select $M_{t} = 32$ antennas or/and $M_{RF} = 16$ RF chains. 
The desirable beam power profile 
 is equal to $1$ at  angle ranges $[-2,2]$ degrees and $[19,23]$ degrees, and is zero otherwise.
The parameter $\alpha$ used in this experiment ranges from $320$ to $1600$, while the learning rate is the same $\beta = 0.02$. 
The beampattern of the designed system is shown in Figure.\ref{fig2}, where one can see that selecting RF chains only and using all antennas performs best, giving rise to the lowest sidelobes. Selecting both RF chains and antennas, (hybrid selection) has the worst performance. %
Since antennas are inexpensive, antenna cost savings are rather insignificant. On the other hand, reduction on RF chains in the HAD array can save more while maintaining good performance.

Running on 12GB memory GPU Titan X maxwell, selection on only antennas or RF chains for the above example took $17$ and $15$ minutes, respectively,  while the selection of  both took $47$ minutes. 

\section{Conclusion}
We have proposed a novel beampattern design framework for MIMO radar by selecting the antennas and RF chains from a mMIMO HAD system. The proposed L2S method leverages softmax neural networks to approximate the selection matrices and optimizes the trainable parameters alternatively. Compared with classification method, the complexity of the softmax selection does not grow exponentially. Numerical results have been provided to validate the performance of the proposed approach, showing that the L2S method is able to achieve the desired beampatterns via selecting a limited number of antennas and RF chains from a dense HAD array. Future work will explore the problem of using L2S to minimize the number of antennas/RF chains subject to an error constraint.

\bibliographystyle{IEEEbib}
\bibliography{ref}

\end{document}